\documentclass[conference]{IEEEtran}
\IEEEoverridecommandlockouts

\usepackage{amsmath,amsfonts}
\usepackage{amsthm}

\usepackage{algpseudocode}
\usepackage{algorithm}

\usepackage{graphicx}
\usepackage{hyperref}
\usepackage{multirow}
\usepackage{soul}
\usepackage{subcaption}
\usepackage{tabularx}
\usepackage{textcomp}
\usepackage{xcolor}
\usepackage{xspace}
\usepackage{multirow}
\usepackage{booktabs}
\usepackage{graphicx}
\usepackage{threeparttable}  
\usepackage{amsmath}
\usepackage{amsfonts}
\usepackage{tcolorbox}
\usepackage{stmaryrd}

\usepackage[normalem]{ulem}
\usepackage{enumitem}

\newtheorem{definition}{Definition}

\definecolor{mygreen}{RGB}{112, 173, 71}
\definecolor{myred}{RGB}{192, 0, 0}
\definecolor{kellygreen}{rgb}{0.3, 0.73, 0.09}

\newcommand{\cut}[1]{}

\usepackage{threeparttable}  
\usepackage{booktabs}
\usepackage{multirow}
\usepackage{bm}

\algdef{SE}[DOWHILE]{Do}{doWhile}{\algorithmicdo}[1]{\algorithmicwhile\ #1}%

 \usepackage{tikz}
 \usetikzlibrary{plotmarks}
 \usepackage{pgfplots}
 \usetikzlibrary{patterns}
 \pgfplotsset{compat=1.3}
 \usetikzlibrary{backgrounds}

\newcommand{\kasetup}{{\mathtt{KA.setup}}}
\newcommand{\kagen}{{\mathtt{KA.gen}}}
\newcommand{\kaagree}{{\mathtt{KA.agree}}}
\newcommand{\aeenc}{{\mathtt{AE.enc}}}
\newcommand{\aedec}{{\mathtt{AE.dec}}}
\newcommand{\ssshare}{{\mathtt{SS.share}}}
\newcommand{\ssrecon}{{\mathtt{SS.recon}}}
\newcommand{\ssexponentrecon}{{\mathtt{SS.exponentRecon}}}

\newcommand{\micro}{MicroSecAgg\xspace}
\newcommand{\microdl}{MicroSecAgg$_{DL}$\xspace}
\newcommand{\microgdl}{MicroSecAgg$_{gDL}$\xspace}
\newcommand{\microcl}{MicroSecAgg$_{CL}$\xspace}
\newcommand{\flamingo}{Flamingo\xspace}

\newcommand{\eseafl}{e-SeaFL\xspace}

\def\BibTeX{{\rm B\kern-.05em{\sc i\kern-.025em b}\kern-.08em
    T\kern-.1667em\lower.7ex\hbox{E}\kern-.125emX}}

\makeatletter
\newcommand{\linebreakand}{%
    \end{@IEEEauthorhalign}
    \hfill\mbox{}\par
    \mbox{}\hfill\begin{@IEEEauthorhalign}
}
\makeatother

\begin{document}

\title{Uncovering Attacks and Defenses in Secure Aggregation for Federated Deep Learning}

\author{\IEEEauthorblockN{Yiwei Zhang}
\IEEEauthorblockA{
\textit{Department of Computer Science} \\
\textit{Purdue University}\\
West Lafayette, USA \\
yiweizhang@purdue.edu}
\and 
\IEEEauthorblockN{Rouzbeh Behnia}
\IEEEauthorblockA{
\textit{School of Information Systems and} \\
\textit{Management} \\
\textit{University of South Florida}\\
Sarasota, USA \\
behnia@usf.edu}
\and
\IEEEauthorblockN{Attila A. Yavuz}
\IEEEauthorblockA{
\textit{Department of Computer Science and} \\
\textit{Engineering} \\
\textit{University of South Florida}\\
Tampa, USA \\
attilaayavuz@usf.edu}
\linebreakand
\IEEEauthorblockN{Reza Ebrahimi}
\IEEEauthorblockA{
\textit{School of Information Systems and} \\
\textit{Management} \\
\textit{University of South Florida}\\
Tampa, USA \\
ebrahimim@usf.edu}
\and
\IEEEauthorblockN{Elisa Bertino}
\IEEEauthorblockA{
\textit{Department of Computer Science} \\
\textit{Purdue University}\\
West Lafayette, USA \\
bertino@purdue.edu}
}

\maketitle

\begin{abstract}
Federated learning enables the collaborative learning of a global model on diverse data, preserving data locality and eliminating the need to transfer user data to a central server.
However, data privacy remains vulnerable, as attacks can target user training data by exploiting the updates sent by users during each learning iteration.
Secure aggregation protocols are designed to mask/encrypt user updates and enable a central server to aggregate the masked information.
%in these protocols 
MicroSecAgg (PoPETS 2024) proposes a single server secure aggregation protocol that aims to mitigate the high communication complexity of the existing approaches by enabling a one-time setup of the secret to be re-used in multiple training iterations.
In this paper, we identify a security flaw in the MicroSecAgg that undermines its privacy guarantees. We detail the security flaw and our attack, demonstrating how an adversary can exploit predictable masking values to compromise user privacy. 
%We detail the detailed security flaw in \micro protocol and the attack steps, demonstrating how the adversary can exploit predictable masking schemes to compromise user privacy. 
Our findings highlight the critical need for enhanced security measures in secure aggregation protocols, particularly the implementation of dynamic and unpredictable masking strategies. 
We propose potential countermeasures to mitigate these vulnerabilities and ensure robust privacy protection in the secure aggregation frameworks.
% More specifically, following their security model, we show that an adversary can trivially gain access to the difference of updates, which can directly undermine the privacy of the user data. 
% we show that the reuse of the secret material in their aggregation protocol enables the adversary to compute a 
% introduces vulnerabilities that can be exploited by malicious parties. This paper investigates a specific attack model targeting the \micro protocol, where the clients use the same reusable masks to protect the updates at each iteration.
% An adversary can sniff network traffic during the aggregation phase to capture masked updates from clients. By leveraging the collected masked data across multiple iterations, the adversary can compute the differences in client updates, thereby inferring sensitive information about the underlying private training data. 
% We detail the detailed security flaw in \micro protocol as well as the attack steps, demonstrating how the adversary can exploit predictable masking schemes to perform member inference attacks, ultimately compromising user privacy. Our findings highlight the critical need for enhanced security measures in secure aggregation protocols, particularly the implementation of dynamic and unpredictable masking strategies. We propose potential countermeasures to mitigate these vulnerabilities and ensure robust privacy protection in the secure aggregation frameworks.
\end{abstract}


\section{Introduction}
% \cite{guo2022microfedml}
% \cite{bonawitz2017practical}
% \cite{guo2022microsecagg}

% MPC

Federated Learning (FL) enables collaborative learning of a shared model between distributed parties while keeping the data local, mitigating data privacy and collection challenges common in traditional centralized learning.  
In large-scale FL, clients with limited computational resources, such as mobile devices, can contribute to training a global model with the assistance of a central server. 
In each iteration, the central server collects the local model updates from clients, trains the global model using the client's local data, aggregates them, and refines them.
However, recent attacks have demonstrated that deploying a plain FL paradigm is insufficient to protect the privacy of the participating users' data~\cite{carlini2021extracting, shokri2017membership, carlini2019secret}. More specifically, these attacks can undermine the confidentiality of the training data by only having access to the user updates. 

To mitigate these risks, secure aggregation protocol has been proposed~\cite{bonawitz2017practical}. 
One prominent approach is FastSecAgg~\cite{kadhe2020fastsecagg}. It utilizes multi-party computation (MPC) techniques to securely aggregate user updates.
% But they tend to suffer from high communication overhead due to the large input sizes inherent in FL.
Another common scheme is masking-based~\cite{behnia2023efficient,bell2020secure,ma2023flamingo}, where random masking terms are added to user updates and finally cancel out during aggregation to prevent disclosure of individual information.
Despite these security advancements, the communication and computational complexity of traditional secure aggregation methods remains a significant challenge, particularly when applied to large-scale FL settings or models such as large language models (LLMs)~\cite{radford2019language} with many participating users.
In response to this, Guo et al. introduced MicroSecAgg (\micro)\cite{guo2022microsecagg}, which improves upon existing methods by employing a one-time setup phase that distributes the necessary secret material for multiple iterations, reducing the overhead caused by continually refreshing the masking terms~\cite{bonawitz2017practical,bell2020secure}.

\vspace{5pt}
\noindent
\textbf{Motivation and Contributions}:
This paper specifically evaluates the \micro protocol and identifies a critical vulnerability that compromises its effectiveness. While \micro introduces significant efficiency improvements, we have discovered that its handling of secret material during the aggregation phase leaves it vulnerable to privacy attacks. 
In particular, an adversary capable of intercepting masked updates can compute differences between user updates, which can be exploited to infer private training data~\cite{carlini2021extracting, shokri2017membership}. 
Such a vulnerability is particularly concerning for applications involving sensitive data, such as healthcare or finance, where data privacy is paramount.

To address this flaw in \micro, we propose several improvements that strengthen the protocol's security while maintaining its efficiency. Specifically, we redesign how secret material is generated and shared to ensure that the masking process remains robust, even in the face of sophisticated inference attacks. We introduce a dynamic masking mechanism that generates unique masks for each iteration, combining a constant shared key with a dynamic component (e.g., a random value or iteration number). This ensures that masked updates remain secure across multiple iterations, preventing adversaries from inferring private data by comparing updates. Furthermore, we propose optimizations to the aggregation phase, which enhance both the security and efficiency of \micro without introducing significant communication or computational overhead.

In summary, our contributions are as follows:
\begin{itemize}
\item We identify a fundamental vulnerability in the \micro protocol that exposes user data to privacy breaches.
\item We present an attack that exploits this vulnerability, demonstrating how an adversary can compromise the privacy of individual users’ training data.
\item We propose improvements to the \micro protocol that enhance its security and efficiency by introducing more robust handling of secret material during the aggregation phase.
\end{itemize}

Through this work, we aim to advance the state of the art in privacy-preserving technologies for federated learning and provide stronger defenses against emerging privacy threats.

\section{Preliminary}
\subsection{Notions and Cryptographic Primitives}

% We use $[n_1, n_2]$ to denote the set of integers $\{n_1, \dots, n_2\}$, and omit the left bound when $n_1 = 1$, i.e., $[n]$ represents the set $\{1, \dots, n\}$.

% Let $p, q$ be primes such that $p = 2q + 1$. A finite field $\mathbb{Z}_p$ consists of the elements $\{0, 1, \dots, p-1\}$, where addition and multiplication are performed modulo $p$.

We use $[n]$ to represent the set $\{1, \dots, n\}$.
Users are denoted by $i$ and Server is denoted by $\mathcal{S}$.
The number of users in a list $\mathcal{U}$ is represented by $|\mathcal{U}|$, while the set of online users is represented by $\mathcal{O}$.
Vectors are denoted by $x_i$, where $i$ 
indexes individual vectors of User $i$. A pseudorandom function $\mathtt{PRF}: \{0,1\}^*\rightarrow r $ is defined where $r$ has the same dimension as the user update vectors $x_i$. 
% In the following, we list the cryptographic building blocks used in \micro protocol. 

\micro employs Shamir’s $t$-out-of-$n$ secret sharing~\cite{shamir1979share}, as defined below, to deal with offline users.
\begin{definition}[Shamir Secret Sharing]   Shamir Secret Sharing allows a secret $s$ to be split into $n$ shares such that any subset of $t$ shares can reconstruct the secret, while fewer than $t$ shares reveal nothing. 
% Given a secret $s \in \mathbb{Z}_q$, the scheme includes the following algorithms:
\begin{itemize}
    \item $\mathtt{SS.share}(s,n,t) \rightarrow \{s_1, \dots, s_n\}$: On the input of a secret $s$, the number of desired shares $n$, and a threshold $t$, it splits the secret $s$ into $n$ shares with a threshold $t$ for reconstruction and returns the $n$ shares $\{s_1, \dots, s_n\}$.
    \item $\mathtt{SS.recon}(\{s_1, \dots, s_t\}, t) = s$: On the input of at least $t$ shares, $\{s_1, \dots, s_t\}$, and the threshold $t$, it reconstructs and return the secret $s$.
\end{itemize}

\end{definition}
% {\color{red} COMMENT: why in the second item above, you say "secret shares"? Before you just used "shares"} 

% An extension, $\mathtt{SS.expoRecon}(\{(g^{s_1}, X_1), \dots, (g^{s_n}, X_n)\}, t) = g^s$, allows reconstruction of the secret in exponent form. Detailed definitions are provided in Appendix A.1.

\begin{definition}[Authenticated Encryption]
Symmetric authenticated encryption (AE) scheme ensures that messages between honest parties cannot be extracted or tampered with by an adversary. The AE scheme is assumed to satisfy IND-CCA2 security, which includes:
\begin{itemize}
    \item $\mathtt{AE.gen}(1^\kappa) \rightarrow k$: On the input of the security parameter $\kappa$, it returns a symmetric key $k$.
    \item $\mathtt{AE.enc}(m, k) \rightarrow c$: On the input of the plaintext $m$ and the key $k$, it encrypts the message and returns the ciphertext $c$.
    \item $\mathtt{AE.dec}(c, k) \rightarrow m$: On the input of the ciphertext $c$ and the key $k$, it decrypts ciphertext and return the plaintext message $m$.
\end{itemize}
\end{definition}

\begin{definition}[Decisional Diffie-Hellman (DDH) Assumption]
Given $p = 2q + 1$, a generator $g$ of $\mathbb{Z}_p^*$, and random values $a, b, c$ chosen from $\mathbb{Z}_q$, the distributions $(g^a, g^b, g^{ab})$ and $(g^a, g^b, g^c)$ are computationally indistinguishable.
\end{definition}

\begin{definition}[Diffie-Hellman (DH) Key Exchange]
 Diffie-Hellman key exchange allows two parties to agree securely on a shared secret over a public channel.
\begin{itemize}
    \item $\mathtt{KA.setup}(\kappa) \rightarrow (G', g, q, H)$: On the input of security parameter $\kappa$, it initializes a group $G'$ of order $q$ with generator $g$ and a cryptographically secure hash function $H$.
    \item $\mathtt{KA.gen}(G', g, q, H) \rightarrow (x, g^x)$: On the input of group $G'$ of order $q$ with generator $g$ and a cryptographically secure hash function $H$, it generates and returns a pair of keys, i.e., a secret key $x$ and corresponding public key $g^x$.
    \item $\mathtt{KA.agree}(x_u, g^{x_v}) \rightarrow s_{u,v} = H((g^{x_v})^{x_u})$: On the input of a secret key $x_u$ of party $u$ and the public key $g^{x_v}$ of party $v$, it computes and returns the shared secret $s_{u,v}$ between two parties $u$ and $v$.
\end{itemize}
\end{definition}

% {\color{blue} Do we use Def 5 in reductions, if not, you can delete it}

% \begin{definition}[DDH Group with an Easy DL Subgroup]
% To handle larger inputs, we adopt the DDH assumption with an easy discrete logarithm (DL) subgroup from \cite{castagnos2015linearly}. This assumption posits a group $G = \langle g \rangle$ with a subgroup $F = \langle f \rangle$, where the DDH problem is hard in $G$ but the DL problem is easy in $F$. 
% % A detailed formal definition is provided in \cite{GroupWithEasyDL}.
% \end{definition}

\subsection{Security Definitions}
In this section, we briefly introduce the definitions of secure aggregation protocol in \micro~\cite{guo2022microsecagg}.

\begin{definition}[Aggregation Protocol]
An aggregation protocol $\Pi(\mathcal{U}, \mathcal{S}, K)$, with a set of users $\mathcal{U}$, a server $\mathcal{S}$, and parameters $K$, consists of two phases: the \textbf{Setup phase} and the \textbf{Aggregation phase}. 
\begin{itemize}
    \item \textbf{Setup phase}: This phase runs once at the beginning of the protocol execution.
    \item \textbf{Aggregation phase}: This phase runs for $K$ iterations. At the beginning of each iteration $k \in [K]$, each user $i \in \mathcal{U}$ holds an input $x_i^k$. At the end of each iteration $k$, the server $\mathcal{S}$ computes and outputs the aggregated value:
    \[
    w_k = \sum_{i \in \mathcal{U}} x_i^k
    \]
\end{itemize}
\end{definition}

\begin{definition}[Correctness with Dropouts]
Suppose the total number of users is $|\mathcal{U}| = n$, the aggregation protocol $\Pi$ is said to ensure correctness with a dropout rate $\delta$ if, for each iteration $1 \leq k \leq K$ and for any set of offline users $\mathtt{offline}_k \subseteq U$ such that $|\mathtt{offline}_k| < \delta n$, the server $\mathcal{S}$ correctly outputs the aggregated value $w_k = \sum_{i\in \mathcal{O} x_i^k}$ at the end of iteration $k$. This holds as long as all users and the server follow the protocol, with the exception that users in $\textit{offline}_k$ may drop offline during iteration $k$.
\end{definition}

% {\color{blue} Do we refer to thee definitions somewhere? If not, eitherdrop them or refer them briefly to harness them.}

\section{MicroSecAgg}
% \rouzbeh{I think it is best to give a high level idea of secure aggregation protocols first, just say how they works etc. then discuss what microsecagg is trying to achieve (efficiency) and how they achieve it (again high-level) - you can get this from their "Our contribution section". 
% Then we would talk about their detailed protocol and talk about their network topology etc.. }
To protect the privacy of a user’s update $w_i$, traditional secure aggregation methods based on multi-party computation (MPC) (e.g., \cite{bonawitz2017practical}) require each user $P_i$ to generate a masking value $h_i$ to obscure their update during aggregation.
% To obscure a client's local update $w_i$, standard MPC-based secure aggregation methods (e.g., \cite{bonawitz2017practical}) require each user $P_i$ to generate a mask $h_i$. 
The mask is constructed so that when all masked updates are aggregated, the sum of the masks cancels out, i.e., $\sum_{i \in [n]} h_i = 0$. To achieve this, each pair of users $P_i$ and $P_j$ need to negotiate pair-wise masks for each iteration -  ensuring that the masks cancel out when both users participate in the aggregation.
Since these masks cannot be reused, users must renegotiate and generate new masks for each aggregation iteration.

However,  traditional methods~\cite{bonawitz2017practical, bell2020secure} suffer from certain drawbacks, requiring users to renegotiate and generate new masks for each aggregation iteration. 
This increases both the communication and computation complexity of such protocols. 
This process involves exchanging information about the freshly generated masks with several others (either all users in \cite{bonawitz2017practical} or a subset in \cite{bell2020secure}), resulting in communication costs growing significantly as the number of users grows.

\micro addresses these limitations with a two-phase secure aggregation protocol, eliminating the need for agreeing on fresh masks at each iteration \cite{guo2022microsecagg}. It introduces reusable masks generated during an initial one-time phase. These masks are then applied consistently across subsequent iterations. Specifically, \micro operates in two distinct phases: the setup phase and the aggregation phase.

% {\color{blue} Is this figure a copy from the MicroSecAgg paper (I did not see it in their figures)? If it is though, we want to mention that the figure is adopted from them in the caption. If we created it, that is fine.}

% \micro leverages MPC, allowing multiple users to collaboratively compute an aggregate result from their private inputs while ensuring that individual inputs remain confidential. 

\begin{figure*}[]
    \centering
    \begin{minipage}{0.49\textwidth}
        \centering
        \includegraphics[width=\textwidth]{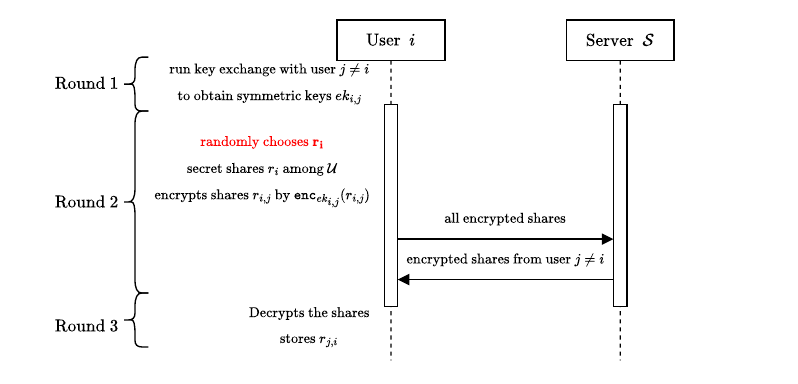}
        \caption*{(a) Setup phase}
    \end{minipage}
    \hfill
    \begin{minipage}{0.49\textwidth}
        \centering
        \includegraphics[width=\textwidth]{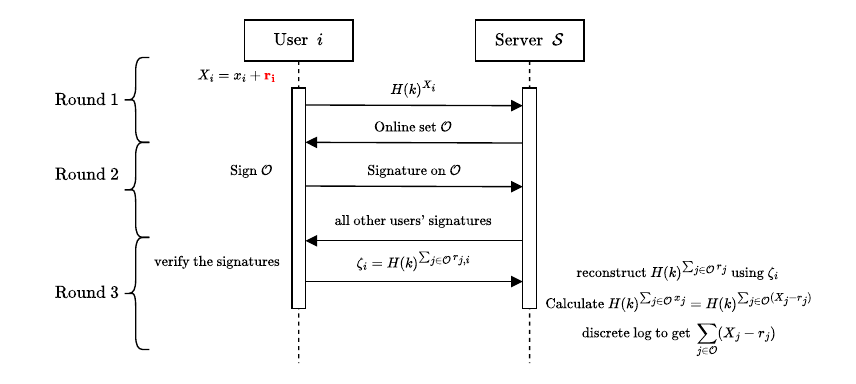}
        \caption*{(b) Aggregation phase}
    \end{minipage}
    \caption{High-level overview of \micro protocol, i.e., \microdl protocol~\cite{guo2022microfedml}.}
    \label{fig:micro}
\end{figure*}
% \rouzbeh{we should mention that this is a MPC-based protocol and what are its building block premitives, eg, secret sharing, ...} 
% To mitigate the communication overhead of negotiating fresh masks for each iteration, 
 
The setup phase, consisting of 3 to 5 rounds (depending on the instantiation), is executed only once when the protocol is initiated. 
During this phase, the server and users generate and exchange their public and private keys. 
The users then employ a secret sharing scheme (i.e., Shamir's Secret Sharing) to create shares of their private inputs and generate masks from these secrets. 
These masks subsequently obscure the local gradients during the aggregation phase.
Users apply the pre-generated masks to their local updates during the aggregation phase, ensuring that individual inputs remain private. 
 Users then send their masked updates to a central server, which aggregates them to compute a global model while preserving the confidentiality of the user data.

They present three instantiations of \micro, namely, \microdl, \microgdl, and \microcl, each designed with different properties, enabling flexibility and adaptability in various applications.
\microdl is the basic version, implementing the core concepts of the two-phase protocol with reusable masks.
\microgdl and \microcl build on the user grouping strategy proposed by Bell {et al.}~\cite{bell2020secure}. In these instantiations, users are divided into groups, and each group runs the \microdl protocol in parallel with the aggregation server.
This grouping strategy reduces communication overhead, as users only interact with a subset of peers while the server aggregates the results by each group.
Additionally, \microcl is optimized for scenarios with larger participant numbers and input sizes, such as update vectors of length $100$ bits. It uses class groups of unknown order, allowing the server to efficiently compute discrete logarithms in the subgroup to recover the sum of updates. In contrast, \microdl and \microgdl handle smaller input domains  (up to 20-bit). 

In the following, we introduce the detailed design of their two-phase secure aggregation protocol. 
In the following, we review \microdl, remarking that other instantiations follow the same main design principles with minor variations. 
Notably, the vulnerability and attack we discuss later on apply to all instantiations of their protocol.
Figure~\ref{fig:micro} shows an overview \micro protocol workflow. Detailed algorithms are listed in Algorithm~\ref{alg:microsetup} and \ref{alg:microagg}.

\subsection{Setup Phase}

\begin{algorithm*}[t]
\caption{\micro~Setup}\label{alg:microsetup}
\small
\raggedright{\textbf{Input:} All parties are provided with the security parameter $\kappa$, the number of users $n$, a threshold value $t$, a Diffie-Hellman key exchange scheme $(\kasetup, \kagen, \kaagree)$, a CCA2-secure authenticated encryption scheme $(\aeenc, \aedec)$, a Shamir's secret sharing scheme $(\ssshare, \ssrecon, \ssexponentrecon)$.}

Every party $i$ holds its only signing key $d_i^{SK}$ and a list of verification keys $d_j^{PK}$ for all other parties.

% \textbf{Input:} A central server $\mathcal{S}$ and a user set $\mathcal{U}$ of $n$ users. Each user can communicate with the server through a private authenticated channel. All parties are given the public parameters: the security parameter $\kappa$, the number of users $n$, a threshold value $t$, honestly generated public parameters $pp \leftarrow \texttt{KA.setup}(\kappa)$ for key agreement, the input space, and a field $\mathbb{Z}_q$ for secret sharing. Moreover, every party $i$ holds its own signing key $d_i^{SK}$ and a list of verification keys $d_j^{PK}$ for all other parties $j$. The server $\mathcal{S}$ also has all users’ verification keys.

\textbf{Output:} Every user $i \in \mathcal{U}$ either obtains a set of users $\mathcal{U}_i$ such that $|\mathcal{U}_i| \geq t$ and a share $r_{j,i}$ of a secret value $r_j$ for each $j \in \mathcal{U}_i$ or aborts. The server either outputs a set of users $\mathcal{U}_S$ such that $|\mathcal{U}_S| \geq t$ or aborts.

\hrulefill 
% \vspace{-1mm}

\textbf{Round 1: Encryption Key Exchange}
% \vspace{-2mm}

% \hrulefill

\begin{algorithmic}[1]
\item Each user $i \in \mathcal{U}$ generates a pair of encryption keys $(sk_i, pk_i) \leftarrow \kagen(pp)$, then signs $pk_i$ with $d_i^{SK}$ and sends $(pk_i, \sigma_i)$ to the server, where $\sigma_i$ denotes the signature.
\item \textbf{Server $\mathcal{S}$:} On receiving $(pk_i, \sigma_i)$ from user $j$, the server verifies the signature $\sigma_i$ with $d_j^{PK}$. If the signature verification fails, it ignores the message from user $j$. Otherwise, it adds $j$ to a user list $\mathcal{U}_S$. If $|\mathcal{U}_S| < t$ after processing all messages from users, the server
%$\mathcal{S}$ 
aborts. Otherwise, the server sends all public keys and signatures it receives from users $j \in \mathcal{U}_S$ to each user in $\mathcal{U}_S$.
\end{algorithmic}

\textbf{Round 2: Mask Sharing}
\begin{algorithmic}[1]
\item Each user $i$: On receiving $(pk_j, \sigma_j)$ for a user $j \in \mathcal{U}$ from the server, each user $i$ verifies the signatures $\sigma_j$ with $d_j^{PK}$. It aborts if any signature verification fails as that indicates the server is corrupt. Otherwise, it inserts $j$ into a user list $\mathcal{U}_i^1$ and stores $ek_{i,j} = \kaagree(pk_j, sk_i)$. It then aborts if $|\mathcal{U}_i^1| < t$ after processing all received messages. Otherwise, user $i$ uniformly randomly chooses $r_i$, and calculates the secret shares of $r_i$ by $r_{i,j} \in \ssshare(r_i, \mathcal{U}_i^1, t)$. Then it encrypts each share $r_{i,j}$ by $c_{i,j} \leftarrow \aeenc(r_{i,j}, ek_{i,j})$ and sends all encrypted shares $\{c_{i,j}\}_{j \in \mathcal{U}_i^1}$ to the server.
\item \textbf{Server $\mathcal{S}$:} If it receives messages from less than $t$ users, it aborts. Otherwise, it denotes this set of users with $\mathcal{U}_S$. It sends each $c_{i,j}$ to the corresponding receiver $j$ for each $i \in \mathcal{U}_S$. Then it outputs the client set $\mathcal{U}_S$.
\end{algorithmic}

\textbf{Round 3: User Receiving Shares}
\begin{algorithmic}[1]
\item Each user $i$: If it receives $c_{j,i}$ for less than $t$ users $j$ from the server, it aborts. Otherwise, it decrypts each encrypted share by $r_{j,i} = \aedec(c_{j,i}, ek_{i,j})$. If the decryption of the share from user $j$ fails, it ignores the encrypted share. Otherwise, it inserts $j$ into a user set $\mathcal{U}_i^2$ and stores $r_{j,i}$. If $|\mathcal{U}_i^2| < t$ after processing all shares, it aborts. Otherwise, it stores $r_i$, the set $\mathcal{U}_i = \mathcal{U}_i^2$, and all $r_{j,i}$ for $j \in \mathcal{U}_i$.
\end{algorithmic}

\end{algorithm*}

Algorithm \ref{alg:microsetup} outlines the setup phase in \micro, which is executed once to initialize the protocol’s keys and configure all entities involved. This phase establishes the cryptographic foundations and sets up  the secure aggregation process.
Specifically, in Round 1, \micro employs a public key infrastructure (PKI) to generate two pairs of private and public keys for all users and the server, which are used for both signature and encryption.
In Round 2, a Diffie-Hellman key exchange enables users to establish shared keys (denoted $ek_{i,j}$) between users $i$ and $j$. Each user then generates a random secret $r_i$, which is used to create a mask for the aggregation phase.
To handle cases where a subset of clients drop out or become unresponsive, \micro employs Shamir’s secret sharing scheme. This allows users to divide their secret $r_i$ into multiple shares ($r_{i,j}$), ensuring that the original secret can be reconstructed only when a sufficient number of shares are combined. This approach not only recovers the secrets of offline users but also enhances privacy by preventing any individual participant from possessing the entire secret.
In Round 3, users distribute their secret shares among others, secured by a CCA2-authenticated encryption scheme using the shared keys $ek_{i,j}$. Upon receiving the encrypted shares, users decrypt and store them for future use.

\begin{algorithm*}[t]
\caption{\micro Aggregation}\label{alg:microagg}
\small
\textbf{Input:} Every user $i$ holds its own signing key $d_i^{SK}$ and all users’ verification key $d_j^{PK}$ for $j \in [n]$, $r_i$, a list of users $\mathcal{U}_i$, and $r_{j,i}$ for every $j \in \mathcal{U}_i$ it obtains in the Setup phase. Moreover, it also holds a secret input $x_i^k$ for every iteration $k$. The server $\mathcal{S}$ holds all users’ verification keys, all public parameters it receives in the Setup phase, and a list of users $\mathcal{U}_S$ which is its output of the Setup phase.

\textbf{Output:} For each iteration $k$, if there are at least $t$ users being always online during iteration $k$, then at the end of iteration $k$, the server $\mathcal{S}$ outputs $\sum_{i \in \mathcal{O}} x_i^k$, in which $\mathcal{O}^k$ denotes a set of users of size at least $t$.

\textbf{Note:} For simplicity of exposition, we omit the superscript $k$ of all variables when it can be easily inferred from the context.

\hrulefill \vspace{-1mm}

\begin{algorithmic}[1]
\For{Iteration $k = 1, 2, \dots$}
    
\hspace{-20pt} \textbf{Round 1: Secret Sharing:}
    \State User $i$: It calculates $X_i = x_i + r_i$ and sends $H(k)^{X_i}$ to the server.
    \State Server $\mathcal{S}$: Denote the set of users it receives messages from with $\mathcal{O}$. If $|\mathcal{O}| < t$, abort. Otherwise, it sends $\mathcal{O}$ to all users $i \in \mathcal{O}$.
    
\hspace{-20pt} \textbf{Round 2: Online Set Checking (Only needed in Malicious setting):}
    \State User $i$: On receiving $\mathcal{O}$ from the server, it first checks that $\mathcal{O} \subseteq \mathcal{U}_i$ and $|\mathcal{O}| \geq t$, then signs the set $\mathcal{O}$ and sends the signature $\sigma_i$ to the server.
    \State Server $\mathcal{S}$: If it receives less than $t$ valid signatures on $\mathcal{O}$, abort. Otherwise, it forwards all valid signatures to all users in $\mathcal{O}$.
    
\hspace{-20pt} \textbf{Round 3: Mask Reconstruction on the Exponent:}
    \State User $i$: On receiving signatures from the server, it first verifies the signatures with $\mathcal{O}$ and the verification keys of the other users. If there are less than $t$ valid signatures, abort. Otherwise, it calculates $\zeta_i = H(k)^{\sum_{j \in \mathcal{O}} r_{j,i}}$. It sends $\zeta_i$ to the server.
    \State Server $\mathcal{S}$: If it receives $\zeta_i$ from less than $t$ users, abort. Otherwise, let $\mathcal{O}'$ denote the set of users $i$ successfully sends $\zeta_i$ to the server. The server reconstructs $R_{\mathcal{O}} = \ssexponentrecon(\{\zeta_j, j\}_{j \in \mathcal{O}'}, t)$ and calculates the discrete log of $H(k)^{\sum_{i \in \mathcal{O}} X_i} / R_{\mathcal{O}}$ to get $\sum_{i \in \mathcal{O}} x_i$.
\EndFor
\end{algorithmic}

\end{algorithm*}

\setlength{\floatsep}{0.1cm}

\subsection{Aggregation Phase}
\micro organizes its aggregation phase into multiple rounds of interaction (i.e., multi-iteration) between the users and the server, allowing for the computation of aggregate results while preserving the privacy of individual updates. Algorithm \ref{alg:microagg} depicts the details of the aggregation phase. 
% Specifically, the Guo \textit{et al.} propose three distinct protocols of \micro framework within this phase: \microdl, \microgdl, and \microcl. 
% Each of these protocols is tailored for different aggregation scenarios, enabling flexibility and adaptability in various applications. 
% But generally, each protocol follows a similar procedure. 
% Specifically, for each iteration, users and the server perform three rounds to ensure securely aggregating the user updates.
In each iteration, the phase unfolds in three distinct rounds, allowing the users and the server to securely aggregate updates.

Initially, in Round 1, each user computes its local gradient update $x_i$ and generates a masked update $X_i$ by combining the gradient update with the secret $r_i$ created during the Setup phase. The user then sends the exponential value $H(k)^{X_i}$ to the server, which maintains a record of all online users and their message.
% Then, the aggregated server will record all the online users.
In Round 2, the server and the users verify the set of online users $\mathcal{O}$ to ensure that at least $t$ users are involved in the 
Aggregation phase.
After that, in Round 3, each user is asked to provide a sum of secret shares of the online users ($\sum_{j\in\mathcal{O}}r_{j,i}$) they store in the Setup phase and to send the exponential value $H(k)^{\sum_{j\in\mathcal{O}}r_{j,i}}$ to the server.
The server finally aggregates the masked updates and secret shares, performs a discrete logarithm to eliminate the masks, and derives the aggregated result without accessing individual updates.
% computes the sum of these masked updates and the secret shares and its discrete log to remove the masks and get the aggregated result without learning about individual updates. 
% \rouzbeh{please explain more about this phase, you can refer to the algorithm 2. Also in their paper they have X=x+r+h not X=x+r only}

\section{\micro Vulnerability}

The reusable masks in \micro significantly enhance aggregation efficiency but introduce a critical vulnerability. Each user employs the same constant mask to protect client updates in every iteration.
As illustrated in Round 1.2 of the Aggregation phase (Algorithm~\ref{alg:microagg}), each user uses $r_i$ as their update masks, while $r_i$ is generated in Round 2.1 of the Setup phase (Algorithm~\ref{alg:microsetup}).
These masks $r_i$ are kept constant and are used repeatedly for each iteration during the Aggregation phase.
% \rouzbeh{here mention what these masks are $r_i$ and $h_i$}. \rouzbeh{We note that based on their design (refer to algo step), the masks are repeated in each iteration} 
If an adversary obtains two masked client updates in two iterations, it can compute the difference between them.
The difference between two gradients will then leak information about the user's training data, potentially compromising privacy~\cite{gong2023gradient,melis2019exploiting}.

% \rouzbeh{you should elaborate that the diff between two gradients will leak info about the user data, use the following refs...}.
% \url{https://link.springer.com/article/10.1007/s10462-023-10550-z} 

% \url{https://link.springer.com/article/10.1007/s10462-023-10550-z}

% This differs from other approaches~\cite{ma2023flamingo, behnia2023efficient}, where distinct masking \rouzbeh{we use the same secret, however we use this secret to derive new masks for each iteration.}terms are used in each iteration. 
This differs from other approaches like \flamingo \cite{ma2023flamingo} and \eseafl \cite{behnia2023efficient}, where although the secret remains unchanged after the setup phase, the masking terms are dynamically derived from the secret using a pseudorandom function (PRF) combined with a seed, such as the iteration number. This ensures that the masking terms differ across iterations, enhancing security.
In contrast, \micro reuses the same secret throughout the aggregation phase, resulting in constant masks across iterations. As a result, this introduces a significant privacy risk, as an adversary could exploit the constant masks to infer sensitive user data.

Note that the \micro implementation of the protocol is only for one gradient update~\cite{micro_git}.
%A special point to note is \micro only presents their protocol for one gradient update in the implementation~\cite{micro_git}.
However, in real world, machine learning models often include at least thousands of gradient elements. 
% Based on the practicality and our engineering experience, 
% {\color{red} COMMENT: the previous sentence is not completed.}
Based on practicality and our engineering experience, each gradient element requires a corresponding mask, indicating the mask length must match the length of the gradient update.
To achieve it, we consider that \micro would generate a new secret for each coordinate of the gradient in the Setup phase and then use it as corresponding mask elements in the Aggregation phase.
In that case, a similar attack can still be applied and expose user training data from the difference of two gradients in two iterations.

\section{Attack on \micro}
% Attacks on FL systems predominantly aim to compromise the integrity and confidentiality of these 
% systems~\cite{bonawitz2017practical,bell2020secure,flamingo}. These include traditional Man-in-the-middle (MITM) attacks and emerging attacks targeting the model to compromise the privacy of the training data.
% In this paper, we propose an attack to compromise \micro by combining both traditional MITM and model inversion.
% In the following, we illustrate our adversary capabilities and the detailed attack steps.

Attacks on federated learning (FL) systems primarily target the integrity and confidentiality of these systems, as outlined in previous work~\cite{bonawitz2017practical,bell2020secure,ma2023flamingo}. These include traditional Man-in-the-Middle (MITM) attacks, as well as more sophisticated approaches like model inversion attacks, which aim to extract sensitive information from the model and its updates. In this paper, we propose a novel attack on \micro that combines both traditional MITM techniques and model inversion methods to compromise the privacy of user training data. In the following, we present the adversary's capabilities and outline the attack steps in detail.

\subsection{Adversary Model}
The primary objective of the adversary in our proposed attack is to undermine the privacy of the user training data. A large body of research~\cite{melis2019exploiting,geiping2020inverting,wang2020sapag,boenisch2023curious,nasr2019comprehensive} has demonstrated that attackers can infer sensitive information about user private data by accessing either the user gradients or their differences (between different iterations).

% In the context of the \micro protocol, 
% \rouzbeh{we should mentioned that micro already considers this adversary in their security model, to claim that our adversary was indeed captured in their security model.}
To launch the attack and compromise the user privacy in \micro protocol, the adversary should be able to intercept and analyze network traffic, capturing the packets exchanged between  the clients and the server during the aggregation phase. 
% Such an adversary capability is already involved int he \micro security model. 
This is feasible given that the communication between clients and the server in FL systems typically occurs over public or semi-secure channels, making them vulnerable to traffic sniffing. 
We note that \micro already accounts for this type of adversary within its security framework. 
% , ensuring that the type of attack we propose falls within its defined threat landscape.
Moreover, the adversary is equipped with sufficient computational power to perform operations such as calculating the discrete logarithm of intercepted messages.
Finally, the adversary is aware of the \micro protocol's structure and has the ability to reverse-engineer or manipulate the updates to infer private information about the data used in local training.

\subsection{Attack Steps}
The attack progresses through the following steps:

\begin{enumerate}
% \item The adversary first sniffs the packets of clients during the setup phase to obtain the mask $r_i$ or $r_{i,j}$.
\item \textbf{Interception of Masked Updates:} During the aggregation phase, the adversary first intercepts the masked updates $H(k)^{X_i}$ sent by users over multiple iterations. 
For instance, the adversary can get gradient updates of user $i$ during iterations $k_1$ and $k_2$ as $H(k_1)^{X_i^{k_1}}$ and $H(k_2)^{X_{i}^{k_2}}$. 
\item \textbf{Computing the Update Differences:} Leveraging the intercepted masked updates, the adversary then calculates the difference between the updates from the user in two different iterations.
This can be done by solving the discrete logarithm problem on the intercepted messages.
By calculating the discrete log of $H(k_1)^{X_{i}^{k_1}}$ and $H(k_2)^{X_{i}^{k_2}}$, the adversary can then get the difference, i.e., $X_{i}^{k_2} - X_{i}^{k_1}$, where $X_{i}^k$ represents the local model update of user $i$ at iteration $i$.
\item \textbf{Model Inversion Attack:} Using the difference between updates across iterations, the adversary performs a model inversion attack, such as gaining private information via the gradient difference~\cite{wang2020sapag}, the gradient inversion attack~\cite{geiping2020inverting}, and the membership inference attack~\cite{nasr2019comprehensive}. 
Typically, this involves inferring information about the training data by exploiting the relationship between the model updates and the underlying data distributions. 
In particular, the adversary attempts to reconstruct private training samples or identify membership information of specific data points.
% The adversary can perform member inference attacks to infer any information about the private training data of client $i$.

\end{enumerate}

Note that for \microgdl and \microcl, an additional masking item is included in the user updates, i.e., $X_i = x_i + r_i + h_i$, where $h_i$ is derived from secrets shared within the user group. 
Similar to $r_i$, $h_i$ is generated during the Setup phase and remains constant throughout the Aggregation phase. 
Therefore, our attack remains effective against \microgdl and \microcl, as the difference in user gradients can be determined using the two masked updates across two iterations.

% Note that for \microgdl and \microcl, they have an additional mask item for the user update, i.e., $X_i = x_i + r_i + h_i$, where $h_i$ is calculated from secrets shared among the user grouping. 
% Similar to $r_i$, $h_i$ is also generate in the Setup phase and keeps constant during the Aggregation phase. Thus, our attack is still efficient for \microgdl and \microcl, as the difference of user gradients can be able to calculated via two masked updates in two iterations.

% \item **Membership Inference Attack:** Based on the information extracted in the previous step, the adversary can conduct a membership inference attack, where they infer whether a specific data point was part of the client’s private training dataset. This is achieved by analyzing how the updates change between iterations and correlating this information with potential training data.
\section{Enhanced Solutions}

% \section{the potential fix - Directions}

In this section, we propose three enhancements to improve \micro secure aggregation protocols, focusing on attack mitigation and efficiency improvements.

\subsection{Addressing the Attack Vulnerability}
The primary cause of the identified attack is the lack of dynamic masking for client updates across iterations.
Currently, \micro negotiates shared secrets between users but uses these secrets directly as static masks without the masks being updated throughout the training process.
As a result, an adversary intercepting traffic can compare updates from the same client across iterations, potentially extracting private training data by calculating the difference~\cite{wang2020sapag}.

To mitigate this, we suggest generating unique masks for each iteration to safeguard client updates. 
Drawing on the approaches in \cite{ma2023flamingo, behnia2023efficient}, clients can create fresh masks for every iteration by combining a constant shared key with a dynamic component, such as a random value known by both parties or the iteration number. 
This dynamic masking would prevent adversaries from inferring private data, even if they obtain masked updates across different iterations.
We note that this can be enabled by adopting methods such as key-homomorphic PRF~\cite{boneh2013key}.
Another method would be to generate $T\times |w|$ shares, where $T$ is the maximum number of iterations and $|w|$ is the size of the weight vector. 
However, the latter would incur significant communication and storage overhead for the setup phase. 
% \micro does not generate a different masks to protect the client updates for each iteration.
% It only negotiates shared secrets for each users and directly uses the secrets as masks without including any variety.
% An intercepting adversary can then sniff the traffics, get the client updates, and calculate the difference of updates of same clients to retrieve user private training data.
% We, therefore, suggest generate different masks to protect client updates for each iteration.
% Inspired by \cite{ma2023flamingo, behnia2023efficient}, clients can generate a fresh masks for each iteration by combining both the constant shared key and a dynamic value. Such a value can be random value known by both shared sides or the iteration number.
% In this way, the adversary will be unable to calculate the update difference even if they can obtain the masked updates for different iteration. 

\subsection{Security Improvements}
Another potential security threat, in addition to the proposed attack, is the vulnerability to model poisoning attacks, which have been shown to affect existing FL protocols~\cite{poison_usenix, poison2, poisonFL}. 
Model poisoning attacks occur when a malicious adversary intentionally manipulates the local model updates to skew the global model’s performance, potentially causing it to misclassify data or behave unpredictably.
These attacks, though not captured by current security models like \cite{guo2022microsecagg}, are especially concerning in critical applications such as malware detection~\cite{malware}, cyber threat intelligence~\cite{threat}, and object detection~\cite{imagenet}, where even minor disruptions can have severe consequences.
% could have severe consequences in critical applications such as malware detection~\cite{malware}, cyber threat intelligence~\cite{threat}, and object detection~\cite{imagenet}.

One major factor enabling these attacks in \micro is the lack of authentication during the transmission of local updates. 
Specifically, in the first round of the aggregation phase (i.e., Secret Sharing), users compute and send their masked updates to the server without any encryption or authentication. An adversary who intercepts these updates could tamper with them, altering or falsifying the values without detection. This could allow a malicious actor to influence the global model’s training, leading to degraded performance or targeted misclassifications.

To mitigate this risk, we propose an enhancement whereby clients sign their masked updates before sending them to the server. By transmitting both the signature and the masked update, the system can ensure the integrity and authenticity of the updates. This measure would prevent adversaries from tampering with the updates, thereby protecting the system from model poisoning attacks.
Given that \micro already uses a key agreement protocol, this signing process can be efficiently implemented using symmetric cryptographic techniques such as HMAC~\cite{bellare1996keying}, which provides both integrity and authenticity with minimal computational overhead. 

% However, with the growing threat of quantum computing, classical cryptographic methods such as HMAC may become vulnerable. To future-proof the security of \micro, it is essential to consider integrating Post-Quantum Cryptography (PQC) techniques to safeguard the system against quantum-based attacks.
% Specifically, post-quantum signature schemes based on lattice-based or hash-based cryptography, such as Dilithium\cite{ducas2018crystals} and SPHINCS+\cite{bernstein2019sphincs+}, offer quantum-resistant alternatives. By incorporating these PQC algorithms, \micro can remain resilient against both classical and quantum adversaries, ensuring that legitimate updates are securely incorporated into the global model even in a post-quantum environment.

\subsection{Efficiency Improvements}
Inspired by \cite{bonawitz2017practical, bell2020secure}, \micro~\cite{guo2022microsecagg} addresses the costly requirement to compute new masking terms at each iteration by utilizing the homomorphic property of Shamir's secret sharing~\cite{shamirSecret}.
They proposed three protocols where \microgdl and \microcl employ the grouping method similar to that in Bell \textit{et al.}~\cite{bell2020secure}.
During the online phase (i.e., the Aggregation phase), the \microdl, \microgdl and \microcl protocols incur ${O}(|U|)$, ${O}(\log |U|)$ and ${O}(\log |U|)$ communication complexity for users, respectively, where $|U|$ represents the number of participating users.
% MicroFedML requires the aggregation server to compute a discrete logarithm and can only work with small weight updates.
Particularly, in the second round of the aggregation phase in \microdl (i.e., Online Set Checking), each user generates a signature on the list of participants and sends it to the server, which distributes it to all users. 
During the third round, i.e., Mask Reconstruction, each user receives $|U|$ signatures and verifies them to ensure $|U| \geq v$ is met, where $v$ is the threshold for the underlying secret sharing scheme. This process imposes $O(|U|)$ communication and computational overhead on each user, as they must verify all received signatures.

To further reduce the overhead, we propose an enhancement for \micro that minimizes the communication and computation complexity for users in malicious settings. 
Instead of each user verifying $O(|U|)$ individual signatures, we suggest adopting multi-signature schemes~\cite{boneh2018compact}, which enable the server to aggregate all user signatures into a single compact signature. 
This modification reduces the complexity for users to $O(1)$, as they would only need to verify one aggregated signature rather than multiple individual ones, thereby streamlining the verification process and minimizing overhead.

\section{Conclusion}
The \micro protocol is an efficient solution for secure aggregation in federated learning as it reduces the communication overhead through reusable masks and a two-phase design. However, the reuse of static masks introduces a significant security vulnerability, allowing adversaries to infer private data by comparing updates across iterations. 
In this paper, we discuss the root cause of such a security vulnerability as well as the practical attack vector.
Moreover, we also outline potential enhanced solutions to mitigate privacy risks and improve the security and efficiency of \micro secure aggregation protocol, making it more robust for real-world federated learning applications.

\section*{Acknowledgment}
% The authors would like to thank the reviewers for their valuable feedback. 
% The work reported in this paper has been supported by NSF under grant 2112471.
% Rouzbeh Behnia was supported by the USF Sarasota-Manatee campus Office of Research through the Interdisciplinary Research Grant program.

The authors would like to express their gratitude to the reviewers for their insightful comments and constructive feedback, which have greatly contributed to the improvement of this paper. This work reported in this paper has been supported by the National Science Foundation (NSF) under Grant No. 2112471. 
Additionally, Rouzbeh Behnia acknowledges the support from the University of South Florida Sarasota-Manatee campus Office of Research through the Interdisciplinary Research Grant program.

\ifCLASSOPTIONcaptionsoff
  \newpage
\fi

\bibliographystyle{IEEEtran}
\bibliography{refs}

\appendices
\end{document}